# Time-resolved imaging of non-diffusive carrier transport in long-lifetime halide perovskite thin films

*Aravindan Sridharan, Nakita K. Noel, Hyeon Hwang, Soroush Hafezian, Barry P. Rand, Stéphane Kéna-Cohen*[*]

A. Sridharan, H. Hwang, S. Hafezian, Prof. S. Kéna-Cohen
Department of Engineering Physics
Polytechnique Montréal
2900 Édouard-Montpetit Blvd, Montreal, QC H3T 1J4, Canada
E-mail: s.kena-cohen@polymtl.ca

Dr. N. K. Noel[*], Prof. B. P. Rand[†]
[*] Princeton Research Institute for the Science and Technology of Materials
[†] Andlinger Center for Energy and the Environment
Department of Electrical Engineering
Princeton University
Princeton, New Jersey 08544, USA

**Abstract**

Owing to their exceptional semiconducting properties, hybrid inorganic-organic perovskites show great promise as photovoltaic absorbers. In these materials, long-range diffusion of charge carriers allows for most of the photogenerated carriers to contribute to the photovoltaic efficiency. Here, time-resolved photoluminescence (PL) microscopy is used to directly probe ambipolar carrier diffusion and recombination kinetics in hybrid perovskites. This technique is applied to thin films of methylammonium lead tri-iodide (MAPbI$_3$) obtained with two different fabrication routes, methylammonium lead tribromide (MAPbBr$_3$), and an alloy of formamidinium lead tri-iodide (FAPbI$_3$) and methylammonium lead bromide FA$_{0.85}$MA$_{0.15}$Pb(I$_{0.85}$Br$_{0.15}$)$_3$. Average diffusion coefficients in the films leading to the highest device efficiencies and longest lifetimes, i.e., in FA$_{0.85}$MA$_{0.15}$Pb(I$_{0.85}$Br$_{0.15}$)$_3$ and acetonitrile-processed MAPbI$_3$, are found to be several orders of magnitude lower than in the other films. Further examination of the time-dependence shows strong evidence for non-diffusive transport. In particular, acetonitrile-processed MAPbI$_3$ shows distinct diffusion regimes on short and long timescales with an effective diffusion constant varying over 2 orders of magnitude. Our results also highlight the fact that increases in carrier lifetime in this class of materials are not necessarily concomitant with increased diffusion lengths and that the PL quantum efficiency under solar cell operating conditions is a greater indication of material, and ultimately device, quality.

# 1. Introduction

Solar energy, which draws upon our nearly infinite supply of solar irradiation, is widely considered to be one of our most important resources for a sustainable energy future. Significant efforts have recently been dedicated to improving a class of solution-processable, metal-halide perovskite materials that when incorporated into solar cells, can yield certified power conversion efficiencies above 23%[1]. Perovskite solar cells surpass their organic and dye-sensitized predecessors in terms of both performance and material cost.[2–6] The prototypical 3D iodide-based perovskites possess low exciton binding energies, small surface and bulk non-radiative recombination rates, a bandgap close to the ideal Shockley-Queisser value for single junction solar cells, and composition dependent bandgap tunability.[7] The rapid increase in the power conversion efficiencies (PCEs) of perovskite solar cells has been unparalleled compared to other photovoltaic technologies.[8] However, there remain many questions regarding the physicochemical processes and optoelectronic properties leading to such high efficiencies. The nature of the photogenerated species, binding energies, as well as recombination pathways are being continuously re-examined.[9–25]

Long carrier diffusion lengths have been identified as an important aspect enabling high efficiencies in planar heterojunction architectures, but reports differ greatly with respect to reported values. In methylammonium lead iodide (MAPbI$_3$) and MAPbI$_{3-x}$Cl$_x$ thin films, i.e. MAPbI$_3$ processed from a PbCl$_2$ precursor to improve film quality, photoluminescence (PL) quenching and transient absorption measurements have been used to extract respective diffusion lengths of ~100 nm and ~1 μm in the direction perpendicular to the substrate. This corresponds to diffusion coefficients $D = 0.01–0.05$ cm$^2$s$^{-1}$.[26,27] In such measurements, the reduced lifetime in the presence of a (perfectly) quenching interface for one carrier type is fit to the diffusion equation. Alternatively, the diffusion coefficient can be obtained *via* Einstein's relation, using measurements of the carrier mobility, μ. In MAPbI$_3$ single crystals, time-of-flight and transport measurements have been used to extract diffusion lengths exceeding 175 μm and up to 3 cm under low fluence excitation.[28] In other single crystal perovskites, mobilities ranging from μ = 105 to 2000 $cm^2V^{-1}s^{-1}$ have been measured using the Hall effect, corresponding to $D$ = 3–50 $cm^2s^{-1}$.[29,30] In MAPbI$_3$ thin films, time-resolved THz conductivity[31] and microwave conductivity measurements[32] have found mobilities of ~30 $cm^2V^{-1}s^{-1}$,[33] corresponding to $D$ = 0.8 $cm^2s^{-1}$. Optical conductivity measurements tend to give slightly higher values than the other techniques given that they yield the *sum* of both carrier mobilities and probe the motion of charge carriers on the few-nanometer length scale.

Although the direction of most importance to photovoltaic efficiencies is the perpendicular direction, lateral diffusion has been shown to be critical in perovskite films due to the impact of grain boundary recombination on device performance.[34] The study of lateral diffusion also lends itself well to spatially-resolved measurement techniques. For example, spatially-resolved photocurrent measurements have been used to extract diffusion lengths of $L_d$ = 10–28 μm (electrons) and 27–65 μm (holes) in MAPbI$_3$ single crystals and to demonstrate unbalanced charge transport in MAPbBr$_3$.[35] All-optical methods such as four-wave mixing,[36] the related light-induced transient grating technique[37] and transient absorption microscopy[38] have been used to analyze in-plane carrier diffusion on thin films without the need for quenching layers. These measurements have yielded values of $L_D$ ($D$) of 1 μm (2 $cm^2s^{-1}$), 0.1–0.3 μm (0.02–0.07 $cm^2s^{-1}$), and 0.2 μm (0.05–0.08 $cm^2s^{-1}$). These values of $D$ correspond to the ambipolar diffusion coefficient, which under high-level injection is described as $D = 2D_nD_p/D_n + D_p$, where $D_n$ and $D_p$ are the electron and hole diffusion coefficients. The broad spread of values suggests both a need for further study, and for techniques which can unambiguously identify diffusion coefficients under a limited number of assumptions. In this work, we directly image the spreading of charge carriers in both time and space and extract the diffusion length from a fit to the 3-dimensional (3D) diffusion equation. Surprisingly, we find that samples with long, monomolecular lifetimes tend to be characterized by small diffusion coefficients, resulting in negligible increases in the diffusion length as compared to the prototypical MAPbI$_3$. Further examination of acetonitrile-processed MAPbI$_3$ shows that in these films non-diffusive transport occurs, where a very large effective diffusion coefficient can be observed only in the first few nanoseconds, on a scale much shorter than the μs lifetime. In contrast, on a scale comparable to the lifetime, the effective diffusion coefficient is more than 2 orders of magnitude smaller. Further examination of the studied samples suggest that this effect is present in most of the films to a varying extent. We also highlight various effects and assumptions that can lead to artifacts in diffusion length measurements. A notable example is the apparent spatial spreading of charge carriers that can occur in the complete absence of diffusion when monomolecular and bimolecular recombination are both present. This effect is analogous to one recently observed in a study of exciton diffusion in monolayer transition metal dichalcogenides in the presence of exciton-exciton annihilation.[39]

## 2. Experiment

To probe carrier diffusion, we record the spreading of luminescence in both space and time, after excitation with a diffraction-limited pump spot, which is used to generate the initial carrier distribution. The technique is depicted schematically in **Figure 1**a. This approach was originally introduced to study exciton diffusion in tetracene single crystals.[40] We implement a slight variant that uses a streak camera to avoid the need for any moving parts.[39] By fitting the time and space-dependent luminescence to the diffusion equation, we directly obtain the recombination and ambipolar diffusion coefficients of free carriers for a range of hybrid organic-inorganic perovskite thin films.

If only linear recombination is present, in-plane diffusion leads to a time-dependent spreading of the gaussian luminescence width $\sigma(t)$, such that $\sigma^2(t) = \sigma^2(0) + 2Dt$. However, a full solution to the diffusion equation becomes necessary when bimolecular recombination occurs. In this case, the spreading is no longer determined by this simple relationship and apparent spreading occurs even in the case $D = 0$.[39] This is the regime typical of free carrier materials and that where our measurements and all previous in-plane measurements of perovskite thin films have been performed.

Prior to performing measurements with a tightly focused pump, we first estimate the linear and bimolecular recombination coefficients by performing time-resolved measurements with a large excitation spot. Under these conditions, diffusion plays a reduced role which allows us to decouple diffusion from the *apparent* lifetime observed when measuring photoluminescence with a diffraction-limited pump. We then perform diffusion measurements by focusing a $\lambda = 532$ nm laser on the sample surface, through the glass coverslip substrates using an oil-immersion objective to a nearly diffraction-limited spot of width $\sigma = 120$ nm (see Experimental Section for details). In this configuration, a slice through the center of the pump spot is imaged through the same objective onto the streak camera entrance slit. This allows us to obtain the spreading of the PL in both space and time as shown schematically on the right-hand side of Figure 1a. Note that the use of immersion oil also prevents any artificial spreading of the spatial profile due to the re-absorption of photons that are totally internally reflected in the glass. Here, we present results for $MAPbI_3$ (**MA**) processed *via* antisolvent quenching, $MAPbI_3$ processed using an acetonitrile/methylamine compound solvent (**MA-ACN**), the mixed-cation, mixed-halide perovskite $FA_{0.85}MA_{0.15}Pb(I_{0.85}Br_{0.15})_3$ (**FAMA**) and $MAPbBr_3$ (**BR**). Scanning electron micrographs of the studied samples are shown in **Figure 2**. We note that the different processing routes and compositions lead to drastically different surface morphology and apparent grain size. For the samples showing clear grain boundaries, the excitation laser was centered within individual domains to minimize any scattering of the gaussian pump beam.

Figure 1b shows the gaussian half-width (standard deviation) obtained by fitting the photoluminescence (PL) spatial profile to a gaussian profile at each time step. For all of the samples, we observe a clear increase in the PL spot size as a function of time. This increase, however, is due to an interplay between bimolecular recombination and diffusion as will be shown below. The difference between the initial PL spot size and the pump size is due to photon recycling which occurs on the ps timescale.[41] Note that the degree of time-dependent spreading varies significantly amongst the various films. One strength of the streak-camera based technique as compared to pump-probe, which is highlighted in Figure 1b, is its adaptability to a wide range of lifetimes, ranging from sub-ns up to ~10 µs.

## 3. Model and Results

From the data, the diffusion and recombination coefficients can be extracted by simultaneously fitting the entire data set (space and time) to the three-dimensional diffusion equation:

$$\frac{\partial \Delta n(r,z,t)}{\partial t} = G(r,z,t) - D\left[\frac{\partial^2 \Delta n(r,z,t)}{\partial r^2} + \frac{1}{r}\frac{\partial \Delta n(r,z,t)}{\partial r} + \frac{\partial^2 \Delta n(r,z,t)}{\partial z^2}\right] - k_1 \Delta n(r,z,t) - k_2 \Delta n(r,z,t)^2 \quad (1)$$

Here, $G(r,z,t)$ is the carrier generation rate determined from the pump, $D$ is the (isotropic) diffusion constant; and $k_1$ and $k_2$ the monomolecular (Shockley-Read-Hall) and bimolecular (band-to-band) recombination coefficients, respectively. Auger recombination can be safely ignored given the low initial carrier densities in our experiments ($< 5 \times 10^{17}$ cm$^{-3}$). We assume high-level injection, i.e. $\Delta n = \Delta p$, where $\Delta n$ and $\Delta p$ correspond to the excess minority carrier density. This assumption is justified when $\Delta n$ exceeds the majority carrier density $n_0$ (or $p_0$). At the carrier densities relevant for most of our experiments, this is a reasonable assumption. Previously reported values are $p_0 \sim 10^{15}$ cm$^{-3}$ in MAPbI$_3$ and MAPbI$_{3-x}$Cl$_x$ thin films, and values ranging from $p_0 = 10^9$–$10^{12}$ cm$^{-3}$ in MAPbI$_3$ single crystals.[28,30,42] Samples from synthetic routes and compositions which yield longer monomolecular lifetimes (fewer traps), such as MA-ACN and FAMA, are expected to possess even lower majority carrier densities. Alternatively, a quadratic dependence of the PL on the pump power, such as that observed BR, can also be used to demonstrate the validity of the high-level injection assumption.[43] Because of the low exciton binding energy (<16 meV) in the materials under study, any excitonic contribution to the photoluminescence at the excitation densities utilised is considered negligible.[44] The reported diffusion coefficients correspond to the ambipolar *D*, which is dominated by the least diffusing species. Under the assumption of high-level injection, the PL can be calculated from the simulated carrier density using PL $\propto \Delta n^2$ . The dependence of the initial carrier density on the pump profile depends on the relative contributions of monomolecular vs bimolecular recombination. In all samples except BR, bimolecular recombination dominates at initial times, i.e. $k_2 \Delta n(t = 0) > k_1$. Given that all of the samples are fabricated on glass and capped with PMMA thin films, we assume reflecting boundary conditions at the top and bottom interfaces.

The fit obtained by a least-squares minimization of the error between our data and the calculated PL is shown in **Figure 3** for MA at a peak excitation density, $\Delta n(0) \equiv \Delta n(r = 0, z = 0, t = 0) = 2.5 \times 10^{16}$ cm$^{-3}$, and in **Figure S2** for the remaining samples. Figure 3a shows a cross-sectional fit to the time-resolved data taken from the center of the spot, while Figure 3b shows spatially-resolved fits taken at various times. The time dependence and spatial spreading are both well-reproduced by the calculation. The full data sets and fits are shown for all the films as **Videos S1-S5**. The fits show excellent agreement with the model, which confirms that invoking the explicit presence of an exciton population or trap states is unnecessary for explaining the results under our measurement conditions (although they may be present). Some small discrepancies arise in the tails at longer times that appear to be due to deviations between the gaussian initial condition (*t* = 0) for the simulation and the real experimental profile (see Experimental Section). A summary of the diffusion and recombination coefficients obtained for the different samples is given in **Table 1**. The 1D monomolecular diffusion length, defined as $L_D = \sqrt{D/k_1}$, is also indicated. Peak carrier density $\Delta n_0$ for each measurement is given in **Table S1**.

For MA, we obtain a value of $D = (1.25 \pm 0.01) \times 10^{-1}$ cm$^2$s$^{-1}$ and a monomolecular diffusion length of $L_D = (1.68 \pm 0.03)$ µm. Processing perovskite films from an acetonitrile/methylamine compound solvent system leads to solar cell efficiencies of up to 19%.[45] Our calculations indicate that the diffusion coefficient in the MA-ACN sample is $D = (6.9 \pm 0.1) \times 10^{-3}$ cm$^2$s$^{-1}$, which is two orders of magnitude lower than that of MA. The much smaller value $k_1$, however, leads to a comparable monomolecular diffusion length $L_D = (1.73 \pm 0.04)$ µm. The FAMA composition was engineered as a thermodynamically stable substitute to the FAPbI$_3$ α-phase perovskite which was found to transition into a non-perovskite δ-phase at room temperature (**Figure S3**).[46] For FAMA, which is the composition of the highest-reported device efficiencies (~22%),[8] the extracted diffusion coefficient is even smaller, $D = (5.6 \pm 0.1) \times 10^{-4}$ cm$^2$s$^{-1}$ with a corresponding diffusion length of $L_D = (0.78 \pm 0.04)$ µm. We have also performed measurements on BR (MAPbBr$_3$) and our results show a large diffusion coefficient of $D = (5.08 \pm 0.05) \times 10^{-1}$ cm$^2$s$^{-1}$. Despite the high value of $D$, the short diffusion length $L_D = (0.96 \pm 0.04)$ µm obtained for BR is a consequence of its very short lifetime. For MA, which is the most studied film, the recombination coefficients agree with previously reported values,[33] with the caveat that we ignore photon recycling in the transverse direction, but implicitly account for it in the lateral direction. The diffusion coefficient we observe for MA is comparable to that obtained in several previous reports (~0.04 cm$^2$s$^{-1}$).[26,27,38] Our results point to an inverse relation between long lifetimes and $D$ for both MA-ACN and FAMA which suggests weakly mobile, localized minority carriers as the source of the PL in long-lifetime films. When MAPbI$_3$ is processed from the acetonitrile/methylamine solvent, we observe a much longer monomolecular lifetime, suggesting a reduced trap density, but at the same time, a diffusion coefficient that is reduced by one order of magnitude compared to MA. The same trend is observed in FAMA, but with an even longer lifetime and smaller diffusion coefficient.

## 4. Potential Artifacts in Time-Resolved PL Measurements

Two potential artifacts need to be considered when analyzing the kinetics. First, we note that diffusion can strongly modify the time-dependent luminescence, particularly in geometries with a tightly focused pump such as in confocal fluorescence lifetime measurements.[47] **Figure 4**a shows a series of linear (monomolecular) PL transients calculated for increasing values of the diffusion coefficient. Diffusion outside of the excitation spot leads to apparent non-linear behavior, which is simply an artefact due to the spatial dynamics. For this reason, initial conditions for our fits were obtained by first measuring $k_1$ and $k_2$ using measurements performed with a large pump spot (with $D = 0$). These values were then refined using the tightly focused data.

Second, bimolecular recombination leads to an artificial spreading of the PL spatial profile in time. To illustrate this, Figure 4b shows the calculated spreading of the gaussian width of the spatial profile for our samples using recombination coefficients from Table 1, but in the absence of diffusion ($D = 0$). The figure shows a large increase of the gaussian width that is purely due to $k_2$. This is analogous to a recent demonstration that exciton-exciton annihilation leads to increases in the apparent $D$ for excitonic materials.[39] This effect becomes more important for increasing carrier densities. At the densities relevant for our measurements ($10^{16}$-$10^{17}$ cm$^{-3}$), both $D$ and $k_2$ play important roles in dictating the time-dependent spreading. This is always the case under high-level injection for $D$ and $k_2$ values typical of perovskite thin films. As a result, one cannot directly use the gaussian width as a function of time to extract the diffusion constant. Note also that the traditional transient grating decay formula, which has previously been used for measuring in-plane diffusion in perovskite films, has not been properly modified to account for bimolecular recombination.[36,48,49]

**5. Photon Recycling**

Finally, photon recycling also plays an important role in dictating the time-dependent behavior in the directions perpendicular and parallel to the substrate. Recent studies show that in films thicker than the absorption length, the measured $k_2$ can differ from the intrinsic $k_2$ by up to one order of magnitude due to photon recycling in the direction perpendicular to the substrate.[41,42] In the simplest picture, emission and re-absorption in this redistributes the initial carrier density over the entire thickness of the film on a sub-nanosecond time scale. Here, we have only reported the apparent $k_2$, without correcting for this effect. In principle, under high-level injection and when bimolecular recombination dominates, the initial carrier density in-plane should agree with that obtained from the laser profile using $\Delta n(0) \propto \sqrt{P(r,0)}$, where $P(r,t)$ is the laser pump profile. However, the mismatch between the laser ($\sigma_L \approx 120$ nm) and the photoluminescence ($\sigma_{PL,t=0} \approx 300 - 580$ nm) gaussian widths is evident from Figure 1c. To understand the difference in initial widths, we have measured the spreading of the spatial profile of MA-ACN at the highest resolution possible in our system (~50 ps) (Figure S2 - insets g and h). The spreading of the spatial profiles is also shown side by side for the short and long timescales in **Figure 5**. Even at this time resolution, there is a large mismatch between $\sigma_L = 120$ nm and $\sigma_{PL,t=0} = 444$ nm. This is indicative of very fast spreading on a < 50 ps timescale. One can show that this agrees well with the anticipated spreading of ~365 nm from a single photon recycling event (see Supplementary). When exciting at energies high above the bandgap, Guo et al. have recently observed fast spreading of carriers over ~100 nm in only 1.2 ps in MA. Although this effect may contribute to the spreading, the spectral redshift that we observe towards the edge of the PL spot (not shown) and the absence of a strong dependence on the pump wavelength is more indicative of photon recycling. In any case, we implicitly correct for the initial in-plane spreading by using the $t = 0$ PL width to determine the initial carrier density. This leads to much lower initial carrier densities than those that would be obtained directly from the pump size. One could appropriately include photon recycling in the theoretical model or modify the initial transverse density to account for redistribution. However, we have verified that although such changes can affect $k_2$ by up to an order of magnitude, the effect on $D$ is negligible (**Table S2**). The values for MA in Table 1 coincide well with *apparent* values of $k_2$ measured for films of similar thickness although one should be wary of the various assumptions made when comparing different values.[41,42] Note, for example, the importance of correctly calculating the carrier density from the pump profile when measuring the bimolecular recombination coefficient. Under high-level injection and at carrier densities where bimolecular recombination dominates, $\Delta n(0)$ is given by the square root of the pump profile in space (or of the $t = 0$ PL profile) . The spatial width of $\Delta n(0)$ thus differs by a factor $\sqrt{2}$ from that of the pump (or PL).

## 6. Non-Diffusive Transport

For the shorter time window in Figure 5, our modelling gives a diffusion coefficient of $D = (1.559 \pm 0.006)$ cm$^2$s$^{-1}$ for MA-ACN. This high value of $D$, which corresponds to a mobility of ~60 cm$^2$V$^{-1}$s$^{-1}$ is more characteristic of intrinsic values measured on single crystals or over short distances using THz conductivity measurements. Our results thus point to two distinct diffusive regimes in the longer lifetime film. Very fast initial diffusion over a time window nearly 1000 times shorter than the monomolecular lifetime, followed by very slow diffusion on longer timescales. One can systematically investigate this effect for all of the film compositions. **Figure 6** shows best fits to the diffusion coefficient over distinct time slices from the data (dashed lines), with $k_1$ and $k_2$ fixed to the values in Table 1. Although the weighted diffusion coefficients are similar to those obtained using the global fit (Table 1), we observe a clear time-dependent decrease of the diffusion coefficient for all of the films except FAMA. The latter is characterized by a very small diffusion constant (undistinguishable from D~0 cm$^2$s$^{-1}$) for all times other than the one shown in the figure. We suspect that these trends are due to very efficient trapping of the initial rapidly diffusing carriers on defects and at grain boundaries, followed by slower de-trapping/re-trapping. This hypothesis is supported by recent measurements of the microscopic surface potential that have found very large potential fluctuations for MA-ACN (~80 meV) as compared to antisolvent-quenched MA (~20 meV).[50] Ščajev *et al.* have also observed a strong dependence of *D* on the initial carrier density and suggested the presence of band-like and localized diffusion mechanisms. The latter was found to be dominant in mixed-halide films and was attributed to a diffusion mechanism that is dominated by carrier hopping and delocalization in localized states due to alloy disorder.[37] Localization due to alloy disorder in the mixed-halide, mixed-cation FAMA film might thus be another source of the reduced *D*.

## 8. Relevance to Photovoltaic Efficiencies

The diffusion length and the photoluminescence quantum efficiency (PLQE) as a function of the photogenerated carrier density, can also be calculated from the recombination coefficients shown in Table 1. In **Figure 7**a, the diffusion length is calculated as a function of the photogenerated carriers via $L_{D,eff}= \sqrt{D/k_{eff}}$, where $k_{eff} = k_1 + k_2 \Delta n$, and as in Table 1, MA, MA-ACN and BR show the highest diffusion lengths. By performing PLQE measurements in an integrating sphere at low carrier density (~10 μW/cm$^2$), we have confirmed that the radiative (low-level injection) component of $k_1$ is negligible (≲ 1%), which further validates the assumption of high-level injection. We thus estimate the PLQE as the ratio of bimolecular recombination to total recombination rate, i.e. $k_2\Delta n/(k_1 + k_2 \Delta n)$. The result shown in Figure 7b ignores Auger recombination and any possible change in the coefficients themselves that can occur due to phase-space filling at high carrier densities.[37] We note that MA-ACN and FAMA show the highest PLQE values at carrier densities typical of solar cell operating conditions (~10$^{15}$ cm$^{-3}$). Given that these film compositions show the best device performance, the low non-radiative loss appears to be the key factor to attaining high-performance optoelectronic devices.[51]

## 8. Conclusions

In summary, we have directly imaged the diffusion of charge carriers in a range of hybrid organic-inorganic perovskite thin films. Our 3D diffusion model shows excellent agreement with experimental data and is used to extract the diffusion and recombination coefficients. Surprisingly, the application of this technique to film compositions that show the best device performance such as $FA_{0.85}MA_{0.15}Pb(I_{0.85}Br_{0.15})_3$ and ACN processed $MAPbI_3$ reveal the smallest diffusion coefficients. Furthermore, we have observed that in the latter, very efficient diffusion does occur, but only on a scale much shorter than the lifetime. Time-resolved fitting of the diffusion coefficient shows strong evidence for non-diffusive transport. In all of the films, except for FAMA, we observe a notable decrease of $D$ as a function of time, on a scale shorter than the carrier lifetime, In general, we find an inverse relationship between carrier lifetime and diffusion coefficient, which may be due to the presence of weakly localized minority carriers in the longest lifetime films. In contrast, we find a strong correlation between the PLQE at carrier densities typical of solar cell operating conditions and reported device performances,[8,45] which indicates that reduced non-radiative loss plays a much greater role than the diffusion length in determining solar cell efficiencies.

**Experimental Section**

*Film fabrication*: For the mixed-cation, mixed halide perovskite, a 1.25 M solution of $FA_{0.85}MA_{0.15}Pb(I_{0.85}Br_{0.15})_3$ (**FAMA**) was prepared using a 4:1 v:v ratio of DMF:DMSO. The solution was spin-coated onto the desired substrate at 1000 rpm for 10 s, followed by 6000 rpm for 35 s. 100 μL of anisole was dropped onto the substrate 35 s after the beginning of the spin-coating. The films were then annealed at 100 °C for 60 min. The same protocol was repeated for the MAPbI$_3$ (**MA**) and MAPbBr$_3$ (**BR**) samples, where the concentration of the precursor solutions was 1 M. In the case of the acetonitrile processed MAPbI$_3$ (**MA-ACN**), a solution of methylamine in ethanol (Sigma Aldrich, 33 wt%) was placed into an aerator which was kept in an ice bath. A carrier gas, N$_2$, was then bubbled into the solution, thus degassing the solution of methylamine. The produced methylamine gas was then passed through a drying tube filled with a desiccant (Drierite and CaO) before being bubbled directly into the acetonitrile (Sigma Aldrich) which contained the perovskite precursors (MAI:PbI$_2$ ratio of 1:1.06 M) at a concentration of 0.5 M. The gas was bubbled into the black dispersion until the perovskite particles were wholly dissolved resulting in a clear, light yellow solution.

*Streak camera measurement*: A schematic of the optical setup alongside explanations is shown in **Figure S4**. The Fianium supercontinuum laser has a tunable repetition rate of 0.1 to 40 MHz with a pulse width of ~48 ps. The repetition rate used for each sample is shown in Table S1. The Hamamatsu C10910 streak camera is equipped with a M10913-11 triggering unit which has a < 20 ps resolution in the smallest time window of 1 ns. The experiments are done in photon counting mode at 10 ms exposure and maximum MCP gain. An electronically delayed signal from the laser system, obtained using a Stanford DG645 delay unit, is used to trigger the streak camera.

*Fitting procedure*: The initial carrier density profile is obtained via a CCD camera image of the pump laser profile. The peak density of excess photogenerated charge carriers $\Delta n_0$ is obtained from:

$$\Delta n_{total} = \int_V \Delta n(r,z,t=0)dV = \Delta n_0 \int_0^{t_s}\int_0^{2\pi}\int_0^{\infty} f(r)g(z)rdrdz \qquad (2)$$

Where $\Delta n_{total} = P_E/E_{ph}$ is the total excess charge carriers obtained via the pulse energy $P_E$ and the photon energy $E_{ph} = hc_0/\lambda_L$. Assuming high level injection. i.e. PL $\propto \Delta n^2$, the horizontal profile of $\Delta n(r, z, t = 0)$, derived from square root of the gaussian fit of the radial photoluminescence profile at t =0, is given by $f(r) = e^{-r^2/4\sigma_{PL,t=0}^2}$, where $\sigma_{PL,t=0}^2$ is the gaussian width. Here, since the gaussian width of the measured spot is much larger than that of the imaging system PSF at the emitted wavelength, we safely ignore its contribution. The profile along z-axis $g(z) = e^{-z/2t_\alpha}$ is given by the square root of the Beer-Lambert equation for light attenuation. Here, $t_\alpha$ is the absorption length at laser wavelength $\lambda_L$, whereas $t_s$ in equation (2) is the physical thickness of the sample. Resolving the integral with the substituted terms results in the following expression for $\Delta n_0$:

$$\Delta n_0 = \frac{\Delta n_{total}}{8\pi\sigma_{PL,t=0}^2 t_\alpha \left(1 - e^{-t_s/2t_\alpha}\right)} \quad (3)$$

Parameters $\lambda_L$, $t_\alpha$, $t_s$, and $\Delta n_0$ are shown in Table S1 for each sample. The obtained $\Delta n_0$ value in equation (3) and spatial profiles presented in equation (2) are subsequently used to fit the 2D (*r-z* axes) data at time *t* = 0. Following the forward Euler discretization scheme, this array is then used as an initial condition to evaluate the subsequent time steps of the diffusion equation in the deconvoluted domain. The resulting 3D array from the simulation, integrated along the z-axis, produces a 2D (*r*, *t*) intensity matrix for the same spatiotemporal coordinates as the streak camera data. The difference between the normalized model and experimental profiles produces an error value for a given set of recombination and diffusion coefficients. The fit coefficients are then found by least squares minimization of the error function.

**Acknowledgements**

Stéphane Kéna-Cohen acknowledges support from the NSERC Discovery Grant Program and the Canada Research Chair in Hybrid and Molecular Photonics. Aravindan Sridharan acknowledges support from NSERC PGS-D as well as Institut d'Énergie Trottier (IET) grant programs. Nakita K. Noel acknowledges funding via a fellowship from the Princeton Center for Complex Materials (PCCM). Barry P. Rand acknowledges support from the Office of Naval Research (ONR) Young Investigator Program (Award #N00014-17-1-2005).

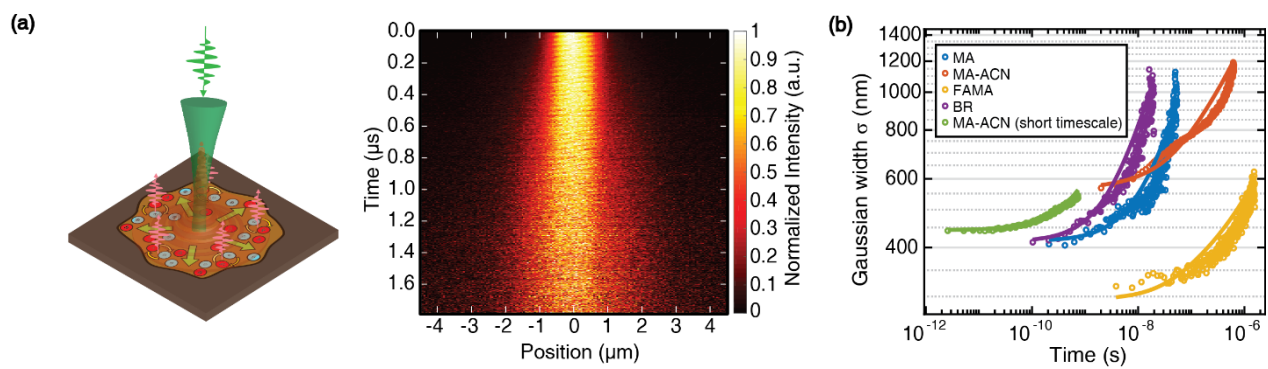

**Figure 1.** a) Left: Schematic of the experiment showing the incoming laser pulse generating electron-hole pairs which diffuse and subsequently recombine, emitting photons that contain information on the spatial distribution of carriers. Right: A streak camera measurement showing spreading of charge carriers in time and space for a film of MAPbI$_3$ processed using an acetonitrile/methylamine (MA-ACN) compound solvent. Data are normalized at each time step to highlight the spreading. b) Log-log plot showing spreading of the standard deviation obtained from a gaussian fit to the spatial profiles as a function of time for samples MA (MAPbI$_3$), MA-ACN on both long and short timescales, FAMA (FA$_{0.85}$MA$_{0.15}$Pb(I$_{0.85}$Br$_{0.15}$)$_3$), and BR (MAPbBr$_3$). The circles and solid lines correspond to raw streak camera data and modelled results respectively.

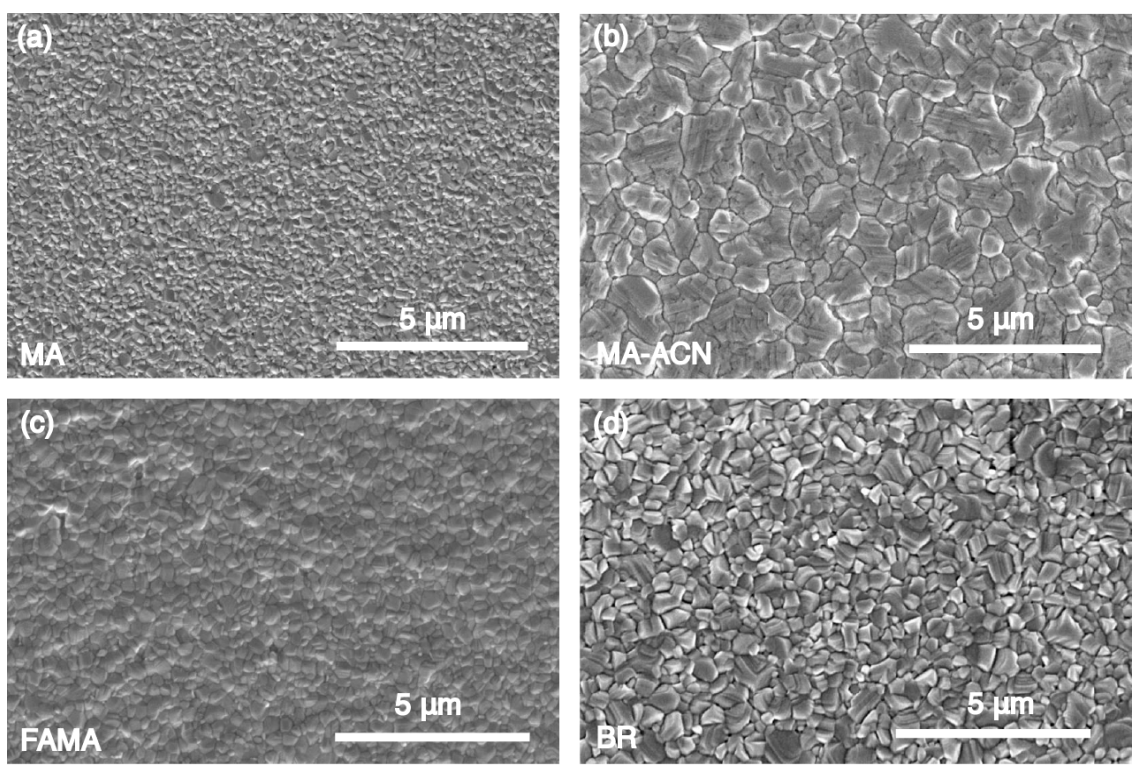

**Figure 2.** SEM images of the studied samples. a) Antisolvent-processed MAPbI$_3$, b) Acetonitrile-processed MAPbI$_3$, c) FA$_{0.85}$MA$_{0.15}$Pb(I$_{0.85}$Br$_{0.15}$)$_3$ and d) MAPbBr$_3$.

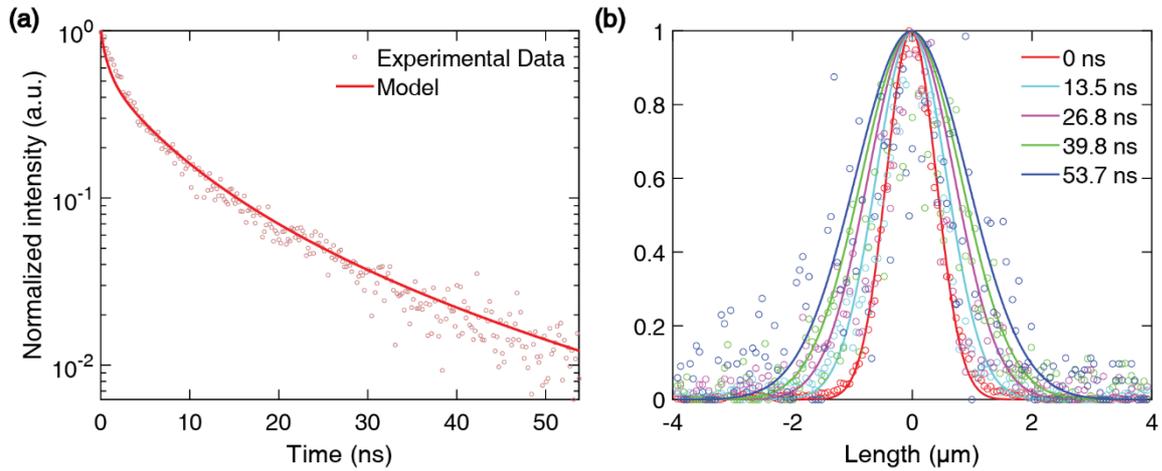

**Figure 3.** a) Photoluminescence transient measured in the center of the spatial profile, i.e. spatial position of 0 µm, and b) Spatial profiles (horizontal slices) for different times extracted from 3D diffusion model-based fitting and experimental data for the MA sample.

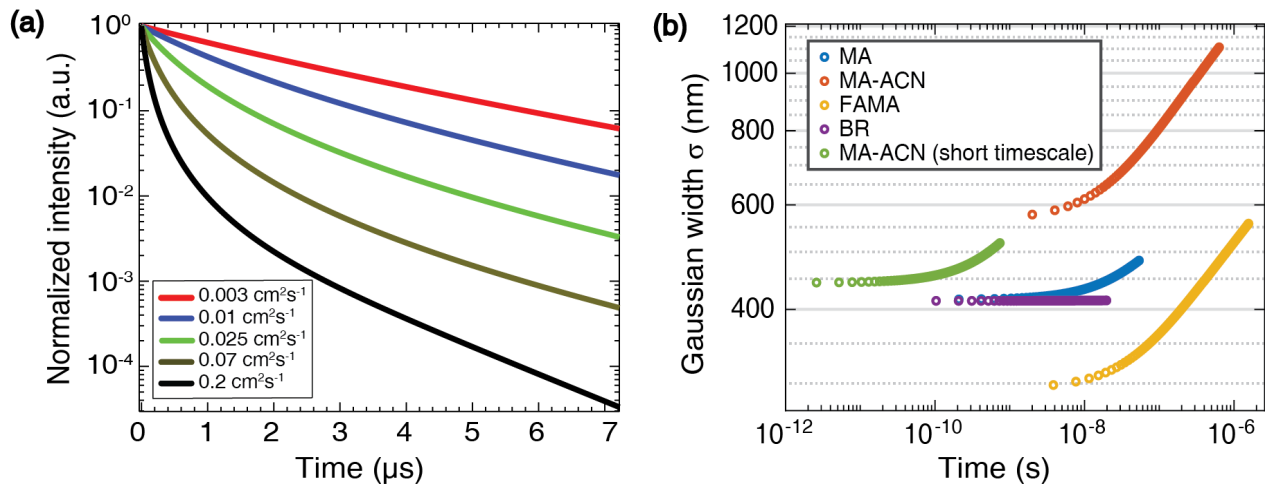

**Figure 4.** a) Impact of the diffusion coefficient on the linear lifetime i.e. with monomolecular recombination only ($k_1 = 1.02 \times 10^5$ s$^{-1}$), in the tightly-focused regime. The rates for bimolecular and Auger recombination are neglected to highlight the significant impact of diffusion alone on the initial part of the lifetime curve. b) Modelled spreading of the gaussian width due to bimolecular recombination coefficient $k_2$ in the absence of diffusion ($D = 0$).

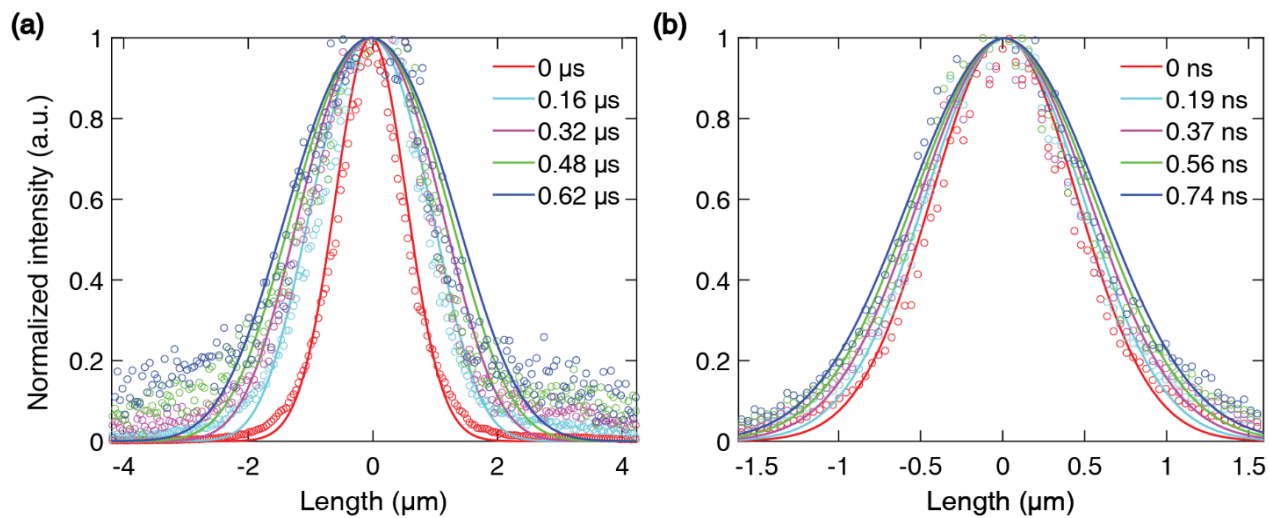

**Figure 5.** Spatial profiles of the PL for the MA-ACN sample at the times indicated in the legend. These are shown a) over a timescale comparable to the monomolecular lifetime and b) at very short times.

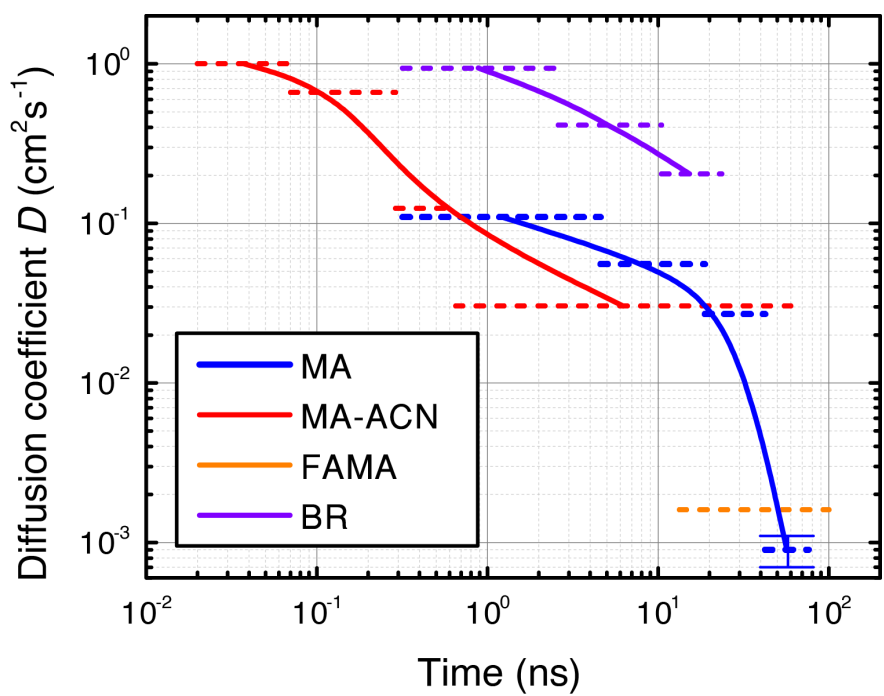

**Figure 6.** Diffusion coefficient, $D$, as a function of time obtained by fitting the data over different time sections, whose ranges are shown as horizontal dashed lines. The full lines are guides to the eye. For MA-ACN the result of the short-time data has been combined with that of the longer time data set. Only one data point is shown for FAMA since $D \approx 0$ cm$^2$s$^{-1}$ reproduces the data very well for all later time sections.

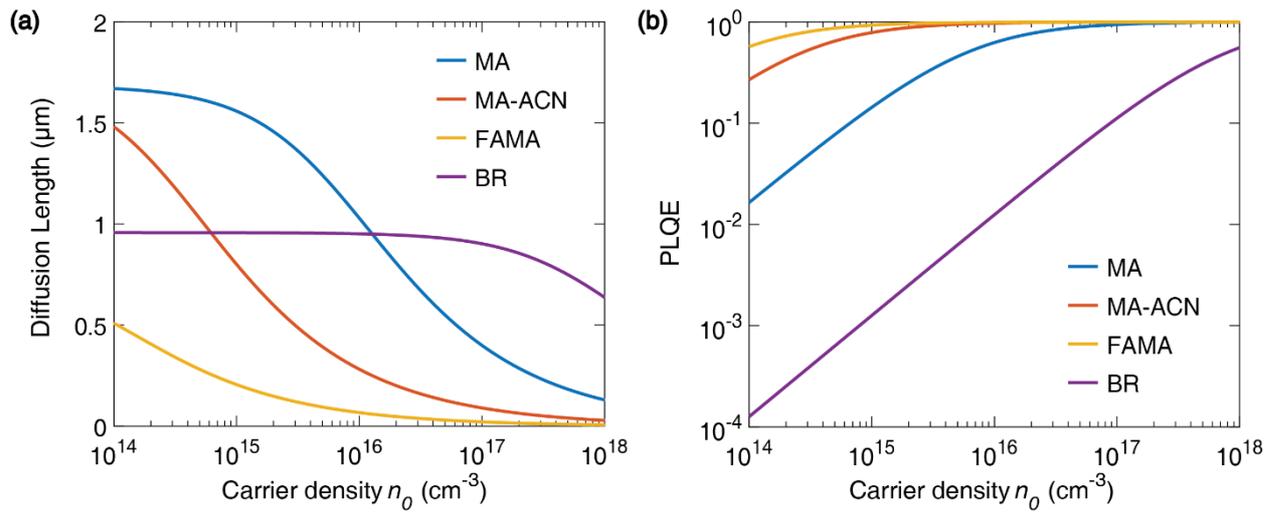

**Figure 7.** a) Diffusion length and b) PLQE as a function of photogenerated carrier density calculated from the coefficients in Table 1 for various metal-halide perovskite films. In both graphs, the lines for BR were calculated using the previously reported value $k_2 = 7 \times 10^{-11}$ cm$^3$s$^{-1}$.[52]

**Table 1.** Diffusion and recombination coefficients obtained through iterative fitting of streak camera data for various perovskite films. Error values correspond to 95% confidence intervals obtained from the fits. The value of $k_2$ is not corrected for photon recycling in the transverse direction and is strongly dependent on the approach used to determine the initial carrier density (see text). For MA, MA-ACN and FAMA, the values of $k_1$ were obtained using separate measurements with larger spot sizes.

| Sample | $D$ [cm$^2$s$^{-1}$] | $k_1$ [s$^{-1}$] | $k_2$ [cm$^3$s$^{-1}$] | $L_D$ [μm] |
|---|---|---|---|---|
| MA | $(1.25 \pm 0.01) \times 10^{-1}$ | $(4.4 \pm 0.1) \times 10^{6}$ | $(7.38 \pm 0.06) \times 10^{-10}$ | $1.68 \pm 0.03$ |
| MA-ACN | $(6.9 \pm 0.1) \times 10^{-3}$ | $(2.331 \pm 0.003) \times 10^{5}$ | $(8.58 \pm 0.08) \times 10^{-10}$ | $1.73 \pm 0.04$ |
| MA-ACN[a] | $(1.56 \pm 0.01)$ | — | — | — |
| FAMA | $(5.6 \pm 0.1) \times 10^{-4}$ | $(9.25 \pm 0.08) \times 10^{4}$ | $(1.23 \pm 0.06) \times 10^{-9}$ | $0.78 \pm 0.01$ |
| BR[b] | $(5.08 \pm 0.05) \times 10^{-1}$ | $(5.55 \pm 0.05) \times 10^{7}$ | — | $0.96 \pm 0.01$ |

[a] Measured in the short 1 ns time window; [b] In BR, $k_2$ could not be obtained due to the dominance of the monomolecular regime $k_1 \gg k_2 n$ under the experimental conditions.

# Supporting Information

**Derivation of gaussian spreading due to a single photon recycling event**

The photon diffusion coefficient is given by:

$$D_\lambda = \frac{c}{3 n_s \alpha_\lambda}$$

Where $c$ is speed of light, $n_s$ is the refractive index of the perovskite film and $\alpha_\lambda$ the absorption at wavelength $\lambda$. The average photon lifetime is given by:

$$\tau_{ph} = \frac{n_s}{c \alpha_\lambda}$$

The average diffusion length of a re-emitted photon prior to reabsorption can then be obtained as:

$$L_{D_\lambda} = \sqrt{D\tau} = \sqrt{\frac{c}{3 n_s \alpha_\lambda} \frac{n_s}{c \alpha_\lambda}} = \sqrt{\frac{1}{3} \frac{1}{\alpha_\lambda}}$$

Taking $\alpha_\lambda = 2.24 \times 10^4$ cm$^{-1}$ for MA-ACN at the peak emission wavelength, i.e. $\lambda = 754.2$ nm (Figure S1), we obtain $L_{D_\lambda} = 2.58 \times 10^{-5}$ cm. We can estimate the temporal spreading of the gaussian width via $\sigma^2(t) = \sigma^2(0) + 2D\tau_{ph} = \sigma^2(0) + 2L_{D_\lambda}^2$, or:

$$\Delta\sigma_{ph.recycling} = \sqrt{\sigma(t)^2 - \sigma(0)^2} = \sqrt{2} L_{D_\lambda}$$

Substituting the above value of $L_{D_\lambda}$, we obtain $\Delta\sigma_{ph.recycling} = 3.65 \times 10^{-5}$ cm or 365 nm.

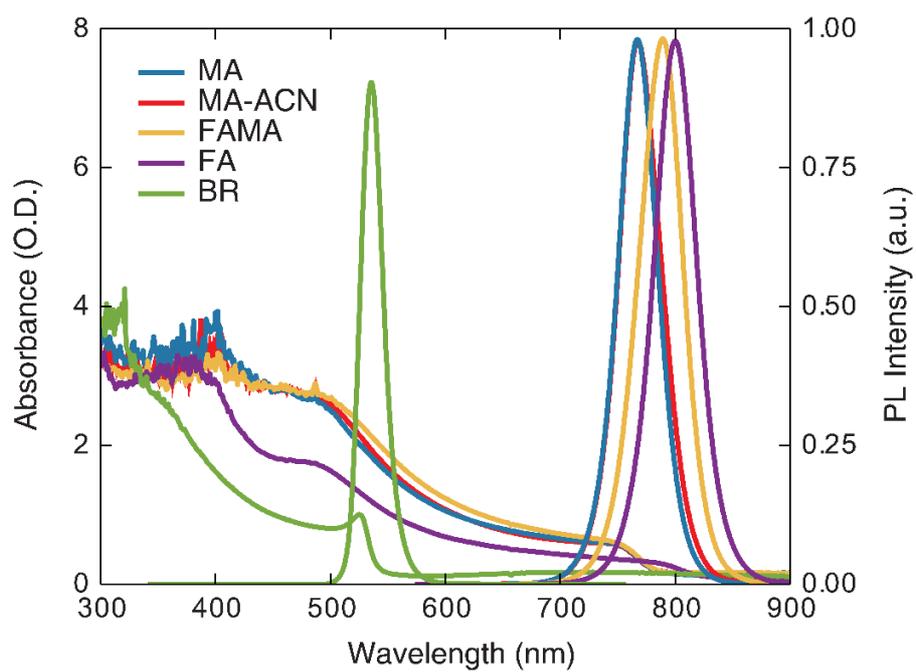

**Figure S1.** Absorption and emission profiles used to calculate the spatial profile at starting time step of simulation.

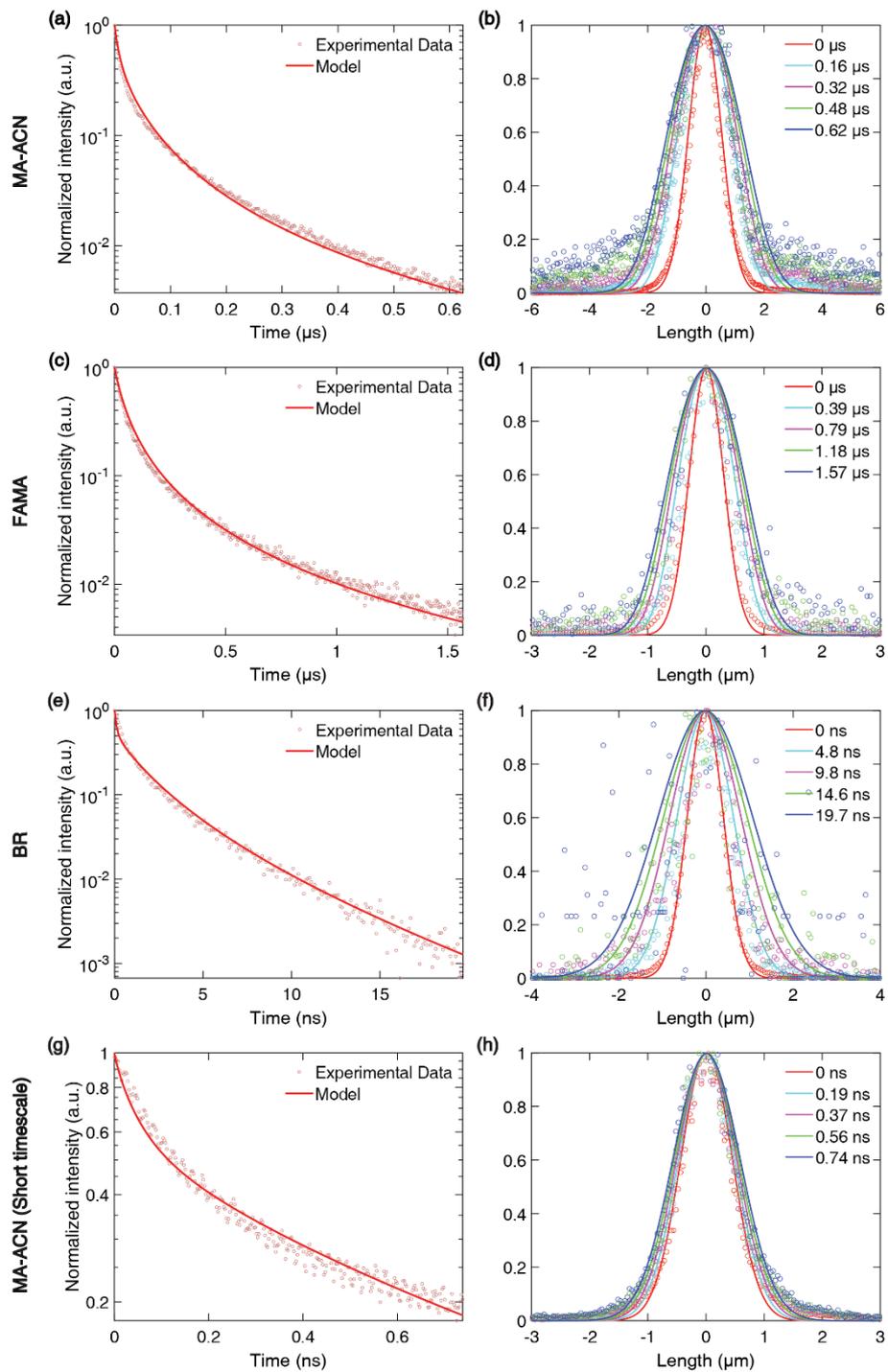

**Figure S2.** Shown for samples MA-ACN (long and short timescales), FAMA, and BR are the time decay profile at the central vertical slice, i.e. spatial position of 0 μm, in a), c), e) and g); spatial profiles (horizontal slices) for different times extracted from 3D diffusion model-based fitting and experimental data in b), d), f) and h).

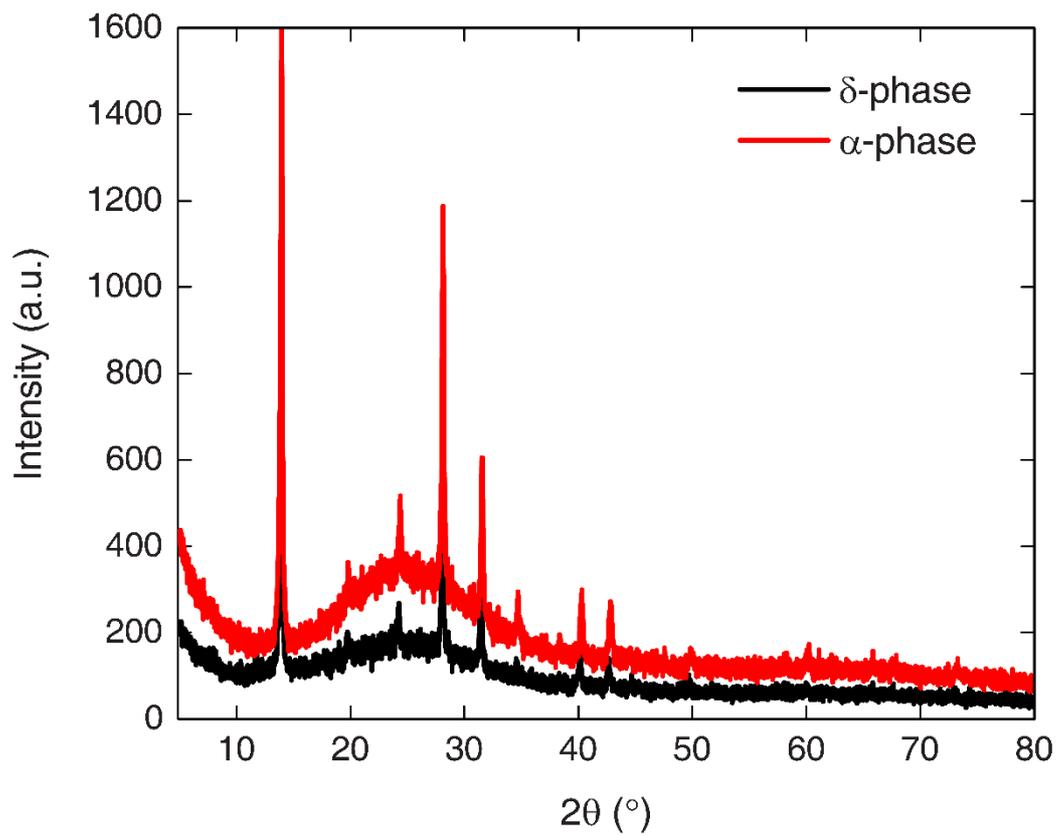

**Figure S3.** XRD data showing FAPbI$_3$ in pure α-phase (experimental condition), and with appearance of δ-phase peaks (near ~12º).

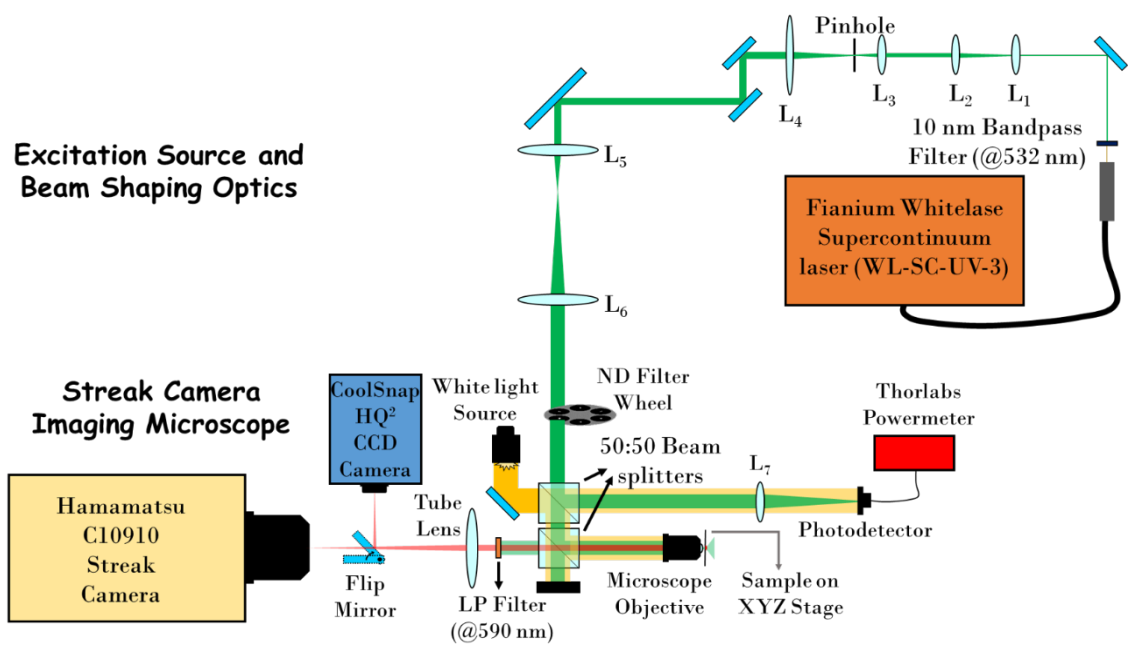

**Figure S4.** Schematic of experimental setup. The optical components for beam shaping are the following: $L_1$ – 150 mm planoconvex lens, $L_2$ – -75 mm planoconcave lens, – 50 mm aspheric lens, $L_4$ – 200 mm biconvex lens, $L_5$ – 100 mm achromatic doublet lens, $L_6$ – 150 mm achromatic doublet lens. $L_1$ and $L_2$ first reducing the beam diameter to half of the original size before focusing onto the pinhole using $L_3$. A 25 μm high-intensity pinhole is used as a spatial filter, which produces a smooth beam profile, between $L_3$ and $L_4$. Subsequently, $L_4$ then re-collimates the beam. Magnification values of 4 and 1.5 are induced by the lens pairs $L_3$-$L_4$ and $L_5$-$L_6$, resulting in a total magnification of 3 from the laser output. Lens $L_7$ is used to focus the light reflected from the first beam splitter onto the power meter. The remaining laser power alongside white light is coupled into the microscopy setup for spatiotemporal imaging through the second beam splitter. An Olympus coverslip-corrected plan apochromatic 60X oil immersion objective of 1.42 NA and 0.15 mm working distance serves the dual purpose of firstly focusing the beam on the sample and secondly, relaying back the collected image from the surface onto the streak/CCD camera through a 200 mm EFL tube lens.

**Table S1.** Experimental conditions and calculated initial parameters for fitting procedure described in experimental section shown for the perovskite thin film samples.

| Sample name | Laser wavelength $\lambda_L$ [nm] | Sample thickness $t_s$ [nm] | Absorption length $t_\alpha$ [nm] | Electron density $n_0$ [$10^{16}$ cm$^{-3}$] | Laser repetition rate (60x objective) [MHz] |
|---|---|---|---|---|---|
| MA | 532 | 398 | 172 | 2.49 | 2 |
| MA-ACN | 532 | 353 | 161 | 4.29 | 0.5 |
| FAMA | 532 | 486 | 192 | 1.25 | 0.1 |
| BR | 450 | 290 | 251 | 0.76 | 20 |

**Table S2.** Simulation results for Sample MA showing fitted coefficient $D$ for varying initial excess carrier densities $\Delta n_0$. These changes in the initial charge carrier densities simulate photon recycling conditions.

| $\Delta n_0$ [$10^{17}$ cm$^{-3}$] | $D$ [$10^{-1}$ cm$^2$s$^{-1}$] |
|---|---|
| 1.5 | $(1.29 \pm 0.01)$ |
| 2.0 | $(1.27 \pm 0.01)$ |
| 2.5 | $(1.25 \pm 0.01)$ |
| 3.0 | $(1.23 \pm 0.01)$ |
| 3.5 | $(1.21 \pm 0.01)$ |